\documentclass[fleqn,usenatbib]{mnras}

\usepackage{newtxtext,newtxmath}

\usepackage[T1]{fontenc}

\DeclareRobustCommand{\VAN}[3]{#2}
\let\VANthebibliography\thebibliography
\def\thebibliography{\DeclareRobustCommand{\VAN}[3]{##3}\VANthebibliography}

\usepackage{graphicx}	
\usepackage{amsmath}	
\usepackage{xcolor}
\graphicspath{{figures/}}

\newcommand{\mystar}{EC\,20187$-$4939}
\newcommand{\e}[1]{$\times 10^{#1}$}

\title[Abundance analysis of a nitrogen-rich hot subdwarf]{Abundance analysis of a nitrogen-rich extreme-helium hot subdwarf from the SALT survey}

\author[L. J. A. Scott et al.]{
L. J. A. Scott$^{1}$\thanks{E-mail: laura.scott@armagh.ac.uk}, C. S. Jeffery$^{1}$, D. Farren$^{2}$, C. Tap$^{2}$, and M. Dorsch$^{3,4}$\\
$^{1}$Armagh Observatory and Planetarium, College Hill, Armagh, BT61 9DB, UK\\
$^{2}$School of Physics, Trinity College Dublin, the University of Dublin, Dublin-2, Ireland\\
$^{3}$Institut für Physik und Astronomie, Universität Potsdam, Haus 28, Karl-Liebknecht-Str. 24/25, 14476 Potsdam-Golm, Germany\\
$^{4}$Dr. Karl Remeis-Observatory \& ECAP, Friedrich-Alexander University Erlangen-Nürnberg, Sternwartstr. 7, 96049 Bamberg,
Germany
}

\date{Accepted 2023 March 1. Received 2023 February 13; in original form 2023 January 17}

\pubyear{2022}

\begin{document}
\label{firstpage}
\pagerange{\pageref{firstpage}--\pageref{lastpage}}
\maketitle

\begin{abstract}
We have performed a detailed spectral analysis of the helium-rich hot subdwarf \mystar{} using data obtained in the SALT survey of helium-rich hot subdwarfs. We have measured its effective temperature, surface gravity and chemical abundances from the spectrum. Its radius has also been determined by fitting the spectral energy distribution using photometric data, from which a mass of 0.44$^{+0.32}_{-0.19}$\,M$_{\odot}$ has been inferred using the measurement of surface gravity. This star is particularly abundant in helium and nitrogen, whilst being both carbon and oxygen-weak. The surface abundances and mass have been found to be consistent with a helium white dwarf merger product. The abundance effects of alpha captures on nitrogen during the merger process and possible connections between \mystar{} and other carbon-weak related objects are discussed.
\end{abstract}




\section{Introduction}
\label{sec:intro}
Hot subdwarfs are evolved helium-burning stars with very thin hydrogen envelopes \citep{heber86,heber16}. Their surface chemical compositions are affected by the abundances of the progenitor star or stars, the formation pathway, and chemical mixing such as diffusion and radiative levitation. As the majority are formed from post-main-sequence stellar cores, hot subdwarfs offer insight into evolution processes which are usually hidden under the envelope. It is therefore valuable to stellar physics in general to understand their composition and formation. The majority of hot subdwarfs have very low-mass hydrogen-rich envelopes and helium-poor surfaces. Approximately 10\%  have helium-rich surfaces with hydrogen being less than 1\% by number \citep{geier13}. Of current interest is what differentiates the hydrogen-deficient from the hydrogen-rich subdwarfs, and whether their evolution pathways can be linked with those of other classes of star. 

Multiple formation pathways for hot subdwarfs are suggested by theory. Binary interaction is thought to be extremely important, if not essential, in the formation process \citep{han02,han03}. In this case, a post-main-sequence star may be stripped by its companion, or the hot subdwarf may be formed by the merging of two white dwarfs. There are also suggested formation pathways involving single stars, although in this case the mass loss mechanism needed to remove the envelope is uncertain \citep{brown01}. Hydrogen-rich surfaces of white dwarfs involved in a merger will be completely mixed with more abundant helium-rich material; thus merger products are more likely to be helium rich. Since helium is produced by the CNO-cycle, enhanced nitrogen abundances  are expected and often seen \citep{stroeer07}. If one of the white dwarfs is sufficiently massive, carbon and possibly oxygen produced during the merger can also be advected to the surface of the product \citep{zhang12a}.  

\mystar{} is a hot subdwarf observed in the Edinburgh-Cape survey of faint blue stars \citep{Cat.EC1} and subsequently in the Southern African Large Telescope (SALT) survey of chemically-peculiar hot subdwarfs \citep{jeffery21a}. It is bright enough to observe at high-resolution (\S\,2) in order to explore connections between cooler extreme helium stars and hot helium-rich subdwarfs via their surface chemistry. For this paper, the spectroscopic data have been analysed to provide temperature, surface gravity and abundance measurements. Photometric data have also been considered and show an excess of flux in the infrared (\S\,3). The results will be discussed with an emphasis on the abundances and the possible formation mechanism of \mystar{} (\S\,4).

\section{Observations}
\label{sec:obs}
\begin{figure*}
	\includegraphics[width=\linewidth]{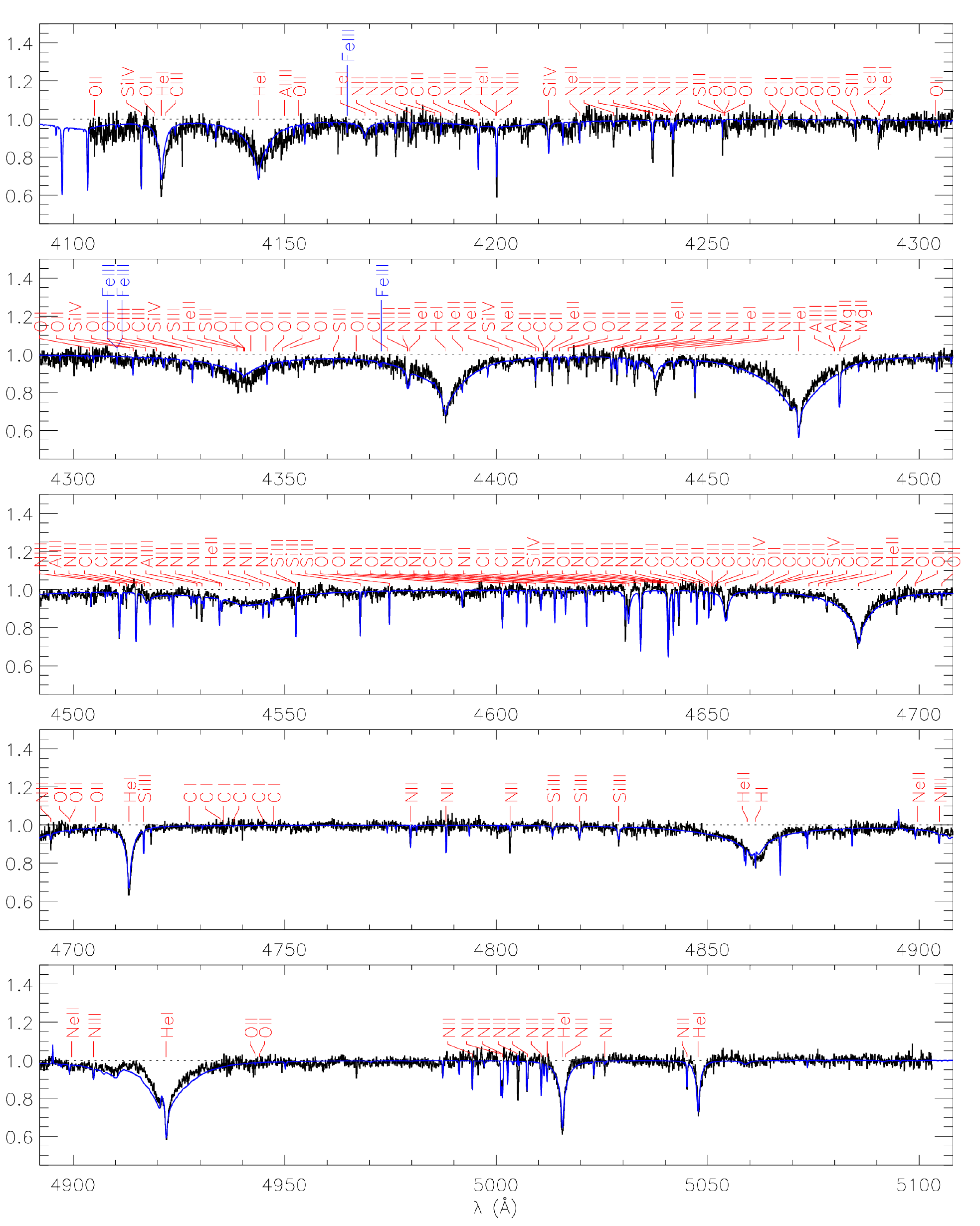}
    \caption{SALT HRS spectrum of \mystar{} in black together with a {\sc tlusty+synspec} model in blue having the parameters of the abundance analysis model (Table~\ref{tab:model_params}) and the abundances listed under $\log{\epsilon_i}$ in Table~\ref{tab:my_abn}. Lines with theoretical equivalent widths $>20$\,m\AA{} are indicated.}
    \label{fig:hrs}
\end{figure*}

The principal spectroscopic data used in this analysis were obtained using SALT's High Resolution Spectrograph (HRS) on 2016 May 25. Two exposures were made in each of the blue (3860--5519\AA) and red (5686--8711\AA) spectral regions. They were reduced to one-dimensional object and sky orders (total counts versus wavelength) using the SALT {\sc pyraf} pipeline \citep{crawford16}. The individual orders were rectified and merged onto a common wavelength grid, and the two observations were combined. The merged blue spectrum is shown in Fig.~\ref{fig:hrs} with spectral lines labelled. The high noise level short-ward of 4100\AA\ makes the blue end unusable. Apart from He{\sc i}, H$\alpha$+He{\sc ii} and many telluric lines, no stellar absorption lines can be identified in the HRS red spectrum, hence only observations in the blue (3860--5519\AA) region have been used in this work. 

In addition, the medium-resolution spectrum of \mystar\ described by \citet{jeffery21a} was also used.   
This spectrum was obtained on 2018 May 17 with SALT's Robert Stobie Spectrograph (RSS).

\begin{table*}
	\centering
	\caption{Equivalent width measurements, $W_{\lambda}$. Lines are sorted by ion and then by wavelength, $\lambda$. For lines measured as part of a blend, wavelengths other than the first of the blend are indicated with a dash. Oscillator strengths as given in the {\sc synspec} linelist are listed under $\log{gf}$.}
	\label{tab:eqw}
	\begin{tabular}{lcl@{\hskip 30pt}lcl@{\hskip 30pt}lcl}
	    \hline
		Ion, $\lambda$ & $\log{gf}$ & $W_{\lambda}$ & Ion, $\lambda$ & $\log{gf}$ & $W_{\lambda}$ & Ion, $\lambda$ & $\log{gf}$ & $W_{\lambda}$\\
		/\,\AA & & /\,m\AA & /\,\AA & & /\,m\AA &  /\,\AA & & /\,m\AA  \\
		\hline
		C\,{\sc III} & & &4523.56 & -0.35 & $37.5 \pm 9.5$ &-~~4430.94 & 0.43 & \\
		4647.42 & 0.07 & $46.2 \pm 12.6$ &4534.57 & -0.43 & $33.2 \pm 9.3$ &-~~4431.00 (Fe\,{\sc III}) & -2.58 & \\
		-~~4647.48 & -2.25 & &-~~4534.52 (Ne\,{\sc II}) & -1.39 & &-~~4431.11 & -0.84 & \\
		-~~4647.43 (Ne\,{\sc II}) & -1.23 & &-~~4534.64 (Ne\,{\sc II}) & -0.67 & && & \\
		-~~4647.66 (Ne\,{\sc II}) & -2.35 & &4535.05 & -0.17 & $25.5 \pm 7.4$ &Mg\,{\sc II} & & \\
		& & &4634.13 & -0.09 & $110.5 \pm 22.7$ &4481.13 & 0.75 & $83.8 \pm 20.1$ \\
		N\,{\sc II} & & &4640.64 & 0.17 & $157.1 \pm 42.6$ &-~~4481.11 & -0.55 & \\
		4124.08 & -0.39 & $17.3 \pm 8.3$ &-~~4641.29 & -2.98 & &-~~4481.32 & 0.59 & \\
		4241.24 & -0.45 & $19.9 \pm 8.8$ &4641.85 & -0.79 & $56.8 \pm 23.4$ && & \\
		4447.03 & 0.22 & $52.8 \pm 13.9$ &-~~4641.96 & -0.29 & &Al\,{\sc III} & & \\
		4601.48 & -0.45 & $52.9 \pm 16.3$ &-~~4641.81 (O\,{\sc II}) & 0.06 & &4529.19 & 0.67 & $25.9 \pm 7.2$ \\
		-~~4601.69 & -0.61 & && & && & \\
		4607.15 & -0.52 & $45.4 \pm 11.5$ &O\,{\sc II} & & &Si\,{\sc III} & & \\
		4613.87 & -0.69 & $43.9 \pm 10.6$ &4649.08 & -4.78 & $30.9 \pm 7.9$ &4567.84 & 0.07 & $61.7 \pm 11.0$ \\
		-~~4613.87 (Ne\,{\sc II}) & -1.50 & &-~~4649.14 & 0.31 & &4574.76 & -0.41 & $29.6 \pm 7.1$ \\
		-~~4613.68 (O\,{\sc II}) & -0.28 & &4661.63 & -0.28 & $11.5 \pm 4.8$ &4716.65 & 0.49 & $36.2 \pm 9.2$ \\
		4621.39 & -0.54 & $40.7 \pm 10.9$ &-~~4661.50 (C\,{\sc II}) & -1.71 & &4813.33 & 0.71 & $27.9 \pm 7.6$ \\
		4779.72 & -0.59 & $40.1 \pm 11.6$ && & &4819.63 & -1.84 & $35.5 \pm 10.0$ \\
		4987.38 & -0.58 & $23.2 \pm 9.0$ &Ne\,{\sc II} & & &-~~4819.71 & 0.82 & \\
		4994.35 & -0.74 & $36.6 \pm 13.9$ &4219.37 & -0.49 & $42.4 \pm 12.9$ &-~~4819.81 & -0.35 & \\
		-~~4994.36 & -0.16 & &-~~4219.74 & 0.71 & &4828.95 & 0.94 & $49.1 \pm 12.4$ \\
		-~~4994.37 & -0.10 & &4290.13 & -4.26 & $72.6 \pm 20.0$ &-~~4829.03 & -3.17 & \\
		5002.70 & -1.03 & $13.5 \pm 5.7$ &-~~4290.37 & 0.92 & &-~~4829.11 & -0.35 & \\
		& & &-~~4290.40 & -0.81 & &-~~4829.21 & -2.17 & \\
		N\,{\sc III} & & &-~~4290.60 & 0.83 & && & \\
		4195.74 & -0.00 & $53.9 \pm 13.0$ &4397.04 & -3.43 & $22.6 \pm 5.2$ &Si\,{\sc IV} & & \\
		-~~4195.97 (N\,{\sc II}) & -0.42 & &-~~4397.99 & 0.63 & &4212.40 & 0.39 & $63.6 \pm 13.7$ \\
		4200.07 & 0.25 & $111.7 \pm 37.0$ &4428.41 & -0.27 & $37.1 \pm 10.3$ &-~~4212.41 & 0.54 & \\
		-~~4199.98 (N\,{\sc II}) & -0.08 & &-~~4428.51 & 0.40 & &-~~4212.41 & -0.76 & \\
		4215.77 & -0.70 & $28.5 \pm 8.9$ &-~~4428.52 & 0.31 & &4328.18 & 0.08 & $34.7 \pm 12.0$ \\
		4510.88 & -0.46 & $67.2 \pm 22.0$ &-~~4428.52 & -1.37 & &4631.27 & 0.96 & $92.9 \pm 20.2$ \\
		-~~4510.96 & -0.59 & &-~~4428.64 & 0.57 & &-~~4631.27 & 0.85 & \\
		4518.14 & -0.46 & $36.8 \pm 15.5$ &4430.90 & 0.21 & $23.5 \pm 7.9$ &-~~4631.27 & -0.58 & \\
		\hline
	\end{tabular}
\end{table*}

\section{Analysis}
\label{sec:analysis}
\mystar{} was included in the coarse analysis of \citet{jeffery21a}, yielding effective temperature $T_{\rm eff}=40.46\pm0.08$\,kK,  surface gravity $\log{g/{\rm cm\,s}^{-2}}=6.02\pm0.04$, and helium-to-hydrogen ratio $\log y = 1.35 \pm 0.13$, from which follow hydrogen and helium number fractions $n_{\rm H}=0.045\pm0.015$ and $n_{\rm He}=0.945\pm0.015$. This work follows on from the coarse analysis in the following stages.

The HRS spectrum was used to measure equivalent widths of identifiable metal lines. The line broadening in the HRS spectrum is <5\,km\,s$^{-1}$, which excludes significant stellar rotation. Following the procedure described by \citet{jeffery17a}, and assuming an LTE model atmosphere with parameters close to those from \citet{jeffery21a}, corresponding abundances were calculated from individual line curves-of-growth using the LTE spectral synthesis code {\sc spectrum} \citep{jeffery01b}.  This provided a first estimate of the surface composition.

These initial abundances were used as input for a grid of non-LTE model atmospheres. This grid was used with the medium-resolution RSS spectral data to obtain new values for $T_{\rm eff}$, $\log{g}$ and $n_{\rm He}$ which consider non-LTE effects on the line profiles. The RSS spectrum was used in order to avoid the profiles of broad lines being affected by the joining of spectral orders, which occurs in the HRS spectrum.

With the atmospheric parameters from the non-LTE grid, a suite of non-LTE model spectra was calculated, varying the abundances of individual metals to create model curves of growth. These were used to measure abundances more accurately by taking into account the effects of other line opacities, as opposed to the single-line profile method used in the first LTE pass.

Finally, publicly available photometric data across a broad wavelength range were fit to model spectral energy distributions (SEDs), with the temperature and surface gravity obtained from the non-LTE grid fit as constraints. This was used to obtain estimates of the mass and radius of \mystar{}. Details of the determination of atmospheric parameters, abundances and SED fitting are given in the following subsections.

\begin{table*}
    \centering
	\caption{Surface abundances of \mystar{} in logarithmic form, $\log{\epsilon_i}$. The number and mass fractions, $n_i$ and $\beta_i$ respectively, are also shown. The subscripts LTE and $\odot$ refer to abundances for the initial LTE estimate and the Sun, respectively. Solar abundances from \citet{asplund09} are given in the usual form of $\log{\epsilon}=\log{n_i/n_{\rm H}}+12$, with the hydrogen abundance defined as $\log{\epsilon_{\rm H}}=12$. Abundances of \mystar{}, both the LTE and final measured abundances, are given as $\log{\epsilon_i}=\log{n_i}+c$, where $c=11.5725$ is a normalisation constant appropriate for He-rich stars. }
	\label{tab:my_abn}
	\begin{tabular}{lccccccc}
		\hline
		Element & $\log{\epsilon_{i,{\mathrm{LTE}}}}$ & $\log{\epsilon_i}$ & $\log{\epsilon_{i,\odot}}$ & $n_i$ & $n_{i,\odot}$ & $\beta_i$ & $\beta_{i,\odot}$\\
		\hline
		H & 10.27 & $9.72\pm0.19$ & 12 & ($1.4\pm0.6$)\e{-2} & $(9.21\pm0.02)$\e{-1} & $(3.53\pm1.51)$\e{-3} & $(7.37\pm0.07)$\e{-1}\\
		He & 11.55 & $11.57\pm0.01$ & $10.93\pm0.01$ & ($9.82\pm0.14$)\e{-1} & ($7.84\pm0.18$)\e{-2} & $(9.85\pm0.14)$\e{-1} & ($2.49\pm0.07$)\e{-1}\\
		C & $8.10\pm0.47$ & $6.94\pm0.26$ & $8.43\pm0.05$ & ($2.33\pm1.40$)\e{-5} & ($2.48\pm0.29$)\e{-4} & $(7.00\pm4.21)$\e{-5} & ($2.36\pm0.34$)\e{-3}\\
		N & $8.24\pm0.04$ & $8.95\pm0.30$ & $7.83\pm0.05$ & ($2.39\pm1.65$)\e{-3} & ($6.22\pm0.72$)\e{-5} & $(8.63\pm5.79)$\e{-3} & ($6.93\pm1.00$)\e{-4}\\
		O & $7.27\pm0.09$ & $7.64\pm0.21$ & $8.69\pm0.05$ & ($1.17\pm0.56$)\e{-4} & ($4.51\pm0.52$)\e{-4} & $(1.69\pm2.24)$\e{-4} & ($5.73\pm0.83$)\e{-3}\\
		Ne & $8.34\pm0.07$ & $8.27\pm0.21$ & $7.93\pm0.10$ & ($4.98\pm2.41$)\e{-4} & ($7.84\pm1.80$)\e{-5} & $(8.76\pm12.2)$\e{-4} & ($1.26\pm0.36$)\e{-3}\\
		Mg & $7.27\pm0.07$ & $7.75\pm0.22$ & $7.60\pm0.04$ & ($1.50\pm0.76$)\e{-4} & ($3.67\pm0.34$)\e{-6} & $(7.28\pm4.62)$\e{-4} & ($7.08\pm0.82$)\e{-4}\\
		Al & $6.06\pm0.06$ & $6.37\pm0.18$ & $6.45\pm0.03$ & ($6.27\pm2.60$)\e{-6} & ($2.59\pm0.18$)\e{-7} & $(1.47\pm1.76)$\e{-5} & ($5.56\pm0.48$)\e{-5}\\
		Si & $7.02\pm0.08$ & $7.52\pm0.19$ & $7.51\pm0.03$ & ($8.86\pm3.88$)\e{-5} & ($2.98\pm0.21$)\e{-6} & $(6.22\pm2.72)$\e{-4} & ($6.65\pm0.58$)\e{-4}\\
		\hline
	\end{tabular}
\end{table*}

\begin{table}
    \centering
    \caption{Model parameters for the {\sc tlusty} grid and the model used for the abundance analysis.}
    \label{tab:model_params}
    \begin{tabular}{l|cc}
        \hline
         & Grid models & Abundance analysis model\\
        \hline
        $T_{\rm{eff}}$ / kK & 39, 40, 41, 42 & 40.78\\
        $\log{g /{\mathrm{cm\,s^{-2}}}}$ & 5.75, 6.00, 6.25 & 5.90\\
        $n_{\rm{H}}$ & 0.1, 0.05, 0.01 & 0.014\\
        \\
        Non-LTE ions & H\,{\sc i-ii}, He\,{\sc i-iii}, & as grid models, plus\\
          & C\,{\sc i-v}, N\,{\sc i-vi}, &  Ne\,{\sc i-v}, Mg\,{\sc ii-iii}\\
          & O\,{\sc i-vi}, Fe\,{\sc ii-vii} & Al\,{\sc ii-iii}, Si\,{\sc ii-v}\\
        \\
        Abundances & Table~\ref{tab:my_abn}, & as grid models, except\\
         & LTE column & $\log{\epsilon_{\rm C}}=6.99$, $\log{\epsilon_{\rm N}}=8.89$, \\
          & & $\log{\epsilon_{\rm{Mg}}}=7.80$, $\log{\epsilon_{\rm{Si}}}=7.57$\\
        \hline
    \end{tabular}
\end{table}

\begin{table}
    \centering
    \caption{Model atom parameters. For each ion, the number of levels and superlevels is given. The next highest ionisation stage for each of the elements listed is treated as a one-level ion.}
    \label{tab:atomic_data}
    \begin{tabular}{lcc@{\hskip 30pt}lcc}
    \hline
    Ion & Levels & Superlevels & Ion & Levels & Superlevels\\
    \hline
    H\,{\sc i} & 8 & 1 & Ne\,{\sc i} & 23 & 12 \\
    He\,{\sc i} & 24 & - & Ne\,{\sc ii} & 23 & 9 \\
    He\,{\sc ii} & 20 & - & Ne\,{\sc iii} & 22 & 12 \\
    C\,{\sc i} & 28 & 12 & Ne\,{\sc iv} & 10 & 2 \\
    C\,{\sc ii} & 17 & 5 & Mg\,{\sc ii} & 21 & 4 \\
    C\,{\sc iii} & 34 & 12 & Al\,{\sc ii} & 20 & 9 \\
    C\,{\sc iv} & 21 & 4 & Si\,{\sc ii} & 36 & 4 \\
    N\,{\sc i} & 27 & 7 & Si\,{\sc iii} & 24 & 6 \\
    N\,{\sc ii} & 32 & 10 & Si\,{\sc iv} & 19 & 4 \\
    N\,{\sc iii} & 25 & 7 & Fe\,{\sc ii} & - & 36 \\
    N\,{\sc iv} & 34 & 14 & Fe\,{\sc iii} & - & 50 \\
    N\,{\sc v} & 10 & 6 & Fe\,{\sc iv} & - & 43 \\
    O\,{\sc i} & 23 & 10 & Fe\,{\sc v} & - & 42 \\
    O\,{\sc ii} & 36 & 12 \\
    O\,{\sc iii} & 28 & 13 \\
    O\,{\sc iv} & 31 & 8 \\
    O\,{\sc v} & 34 & 6 \\

    \hline 
    \end{tabular}
\end{table}

\subsection{Atmospheric parameters}
\label{sec:atm_params}

Stellar atmosphere models and emergent spectra were computed using {\sc tlusty} and {\sc synspec} \citep{hubeny17a,hubeny17b,hubeny17c,hubeny21}. These were then fitted to the data to obtain the atmospheric parameters.

The grid of {\sc tlusty} and {\sc synspec} models ranged from 39 to 42\,kK in temperature (steps of 1\,kK) and from 5.75 to 6.25 in $\log{g/\mathrm{cm\,s^{-2}}}$ (steps of 0.25). The grid included hydrogen number fractions of $n_{\rm H}=0.1$, 0.05 and 0.01. Each model had a microturbulent velocity of 0. Metal abundances were taken from the results of the initial LTE analysis (see Sec.~\ref{sec:abunds}, Table\,\ref{tab:my_abn}), leaving a helium fraction of $n_{\rm He}=1-n_{\rm H}-n_{\rm Z}$, with $n_{\rm Z}$ being the number fraction of metals. The grid models included elements up to atomic number 30. Elements not listed in Table~\ref{tab:my_abn} were assumed to be at solar abundance, with values taken from \citet{asplund09}. Opacity contributions were calculated for H, He, C, N, O and Fe in non-LTE. The grid models are summarised in Table~\ref{tab:model_params}. Information on the model atoms is listed in Tab.~\ref{tab:atomic_data}. The model atoms, which are distributed with version 208 of {\sc tlusty} \citep{hubeny21}, used the NIST database\footnote{https://www.nist.gov/pml/atomic-spectra-database} \citep{ralchenko20} for level energies and the Opacity Project database\footnote{http://cdsweb.u-strasbg.fr/topbase/topbase.html} \citep{cunto93} for photoionization cross sections. For Fe specifically, the Kurucz atomic database\footnote{http://kurucz.harvard.edu/atoms.html} was used.

The medium-resolution data were chosen for the model grid fitting, since the profiles of the broad lines were affected by the joining of spectral orders in the high resolution data. The fitting was performed with the software {\sc sfit} \citep{jeffery01b} using a chi-squared minimisation procedure with extra weighting on the H and He lines.

A range of combinations of $T_{\rm{eff}}$, $\log{g}$ and $n_{\rm{He}}$ were able to provide a similar fit to the data. To break this degeneracy, the abundance ratio of N\,{\sc ii} and N\,{\sc iii} was measured for all temperatures in the grid at $\log{g/\mathrm{cm\,s^{-2}}}=6$. This was measured using the curve of growth method (see Sec~\ref{sec:abunds}). Least squares fitting was then used to find the temperature at which the abundances measured from both ions of N were equal. This temperature was then kept fixed whilst $\log{g}$ and $n_{\rm{He}}$ were varied in the chi-squared minimisation. This gave $T_{\rm eff}=40.78\pm0.35$\,kK, $\log{g/{\mathrm{cm\,s^{-2}}}}=5.90\pm0.20$ and $n_{\rm H}=0.014\pm0.006$.

\begin{figure*}
    \centering
    \includegraphics[width=\columnwidth]{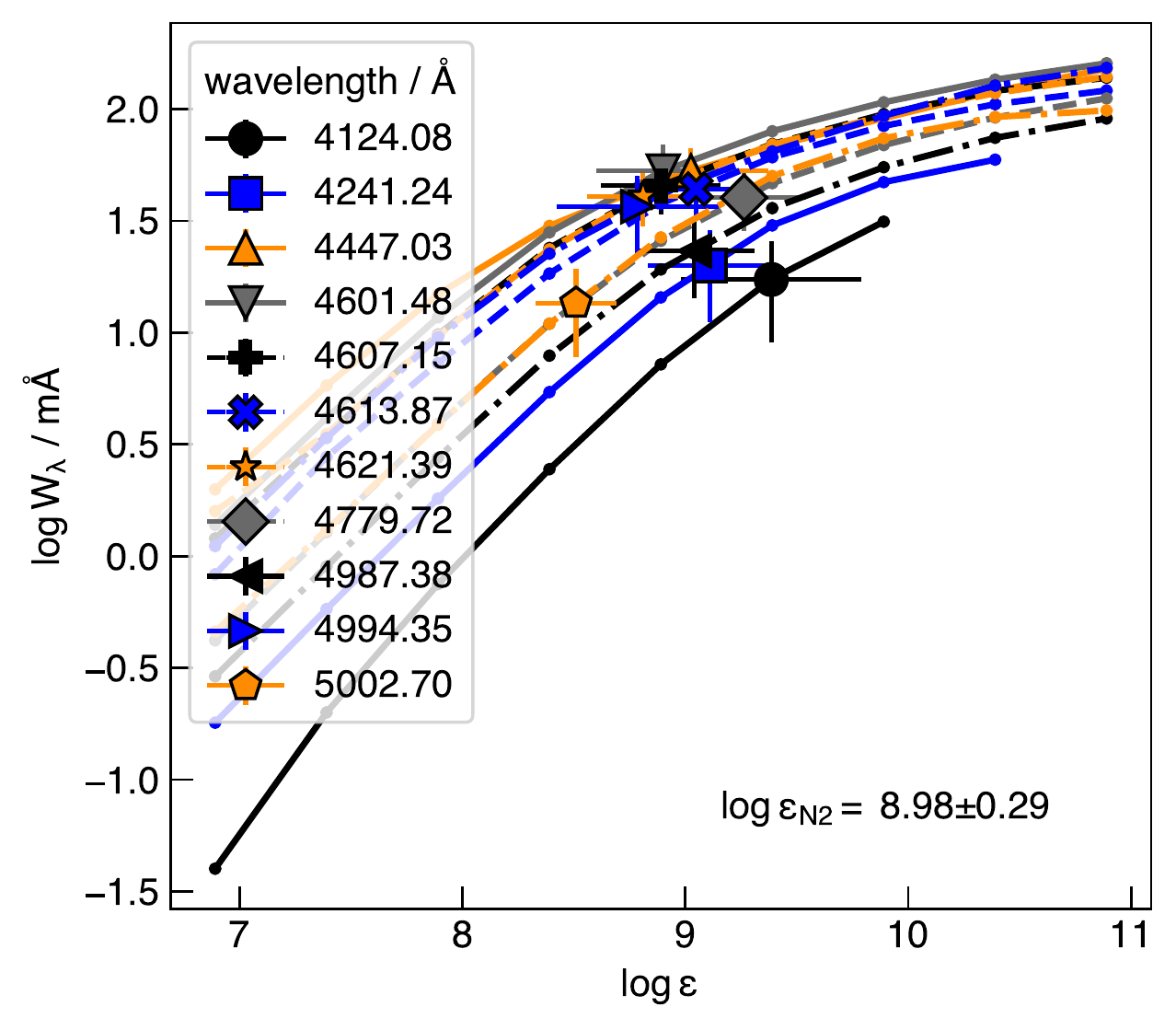} \includegraphics[width=\columnwidth]{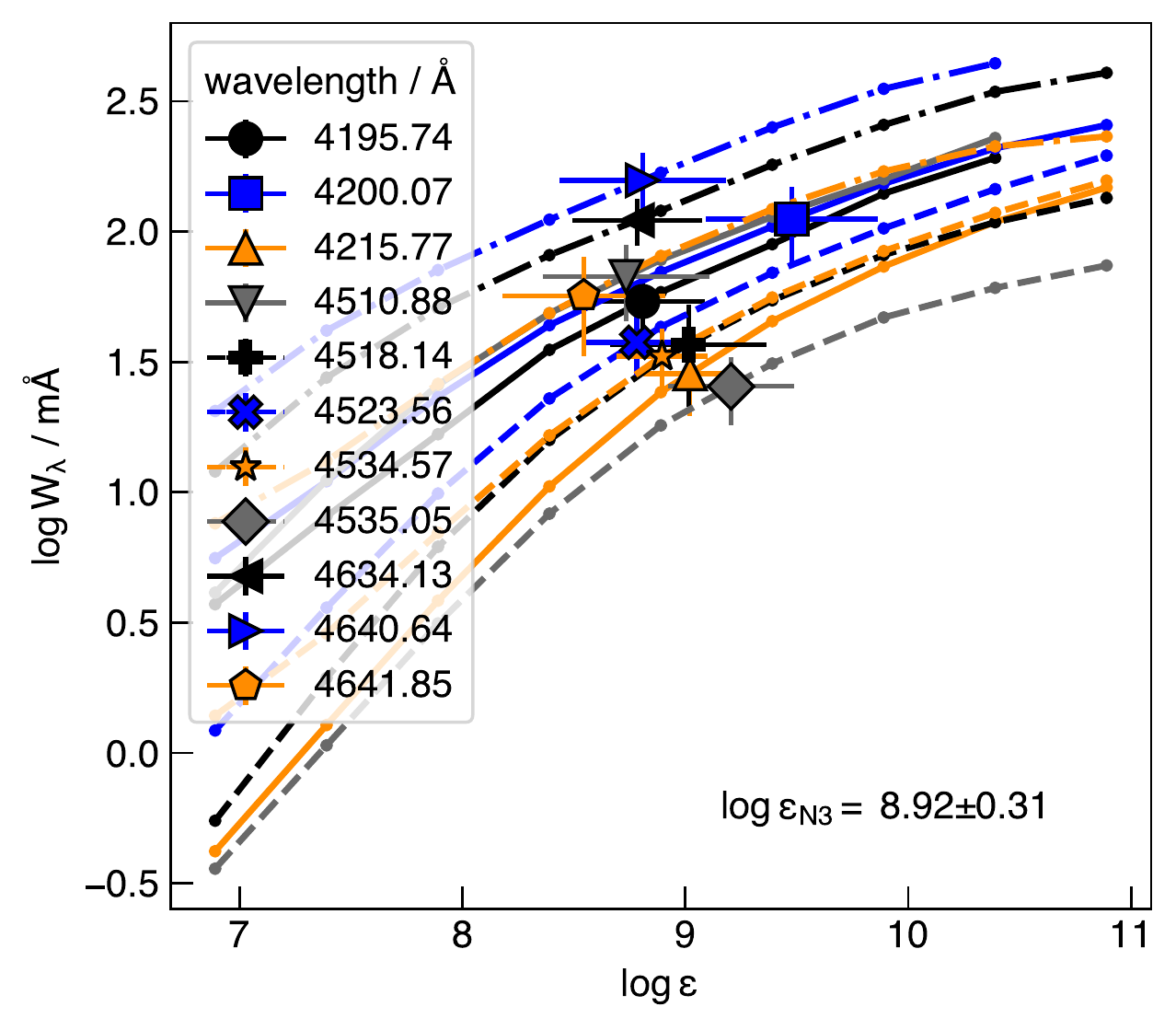} \\
    \includegraphics[width=\columnwidth]{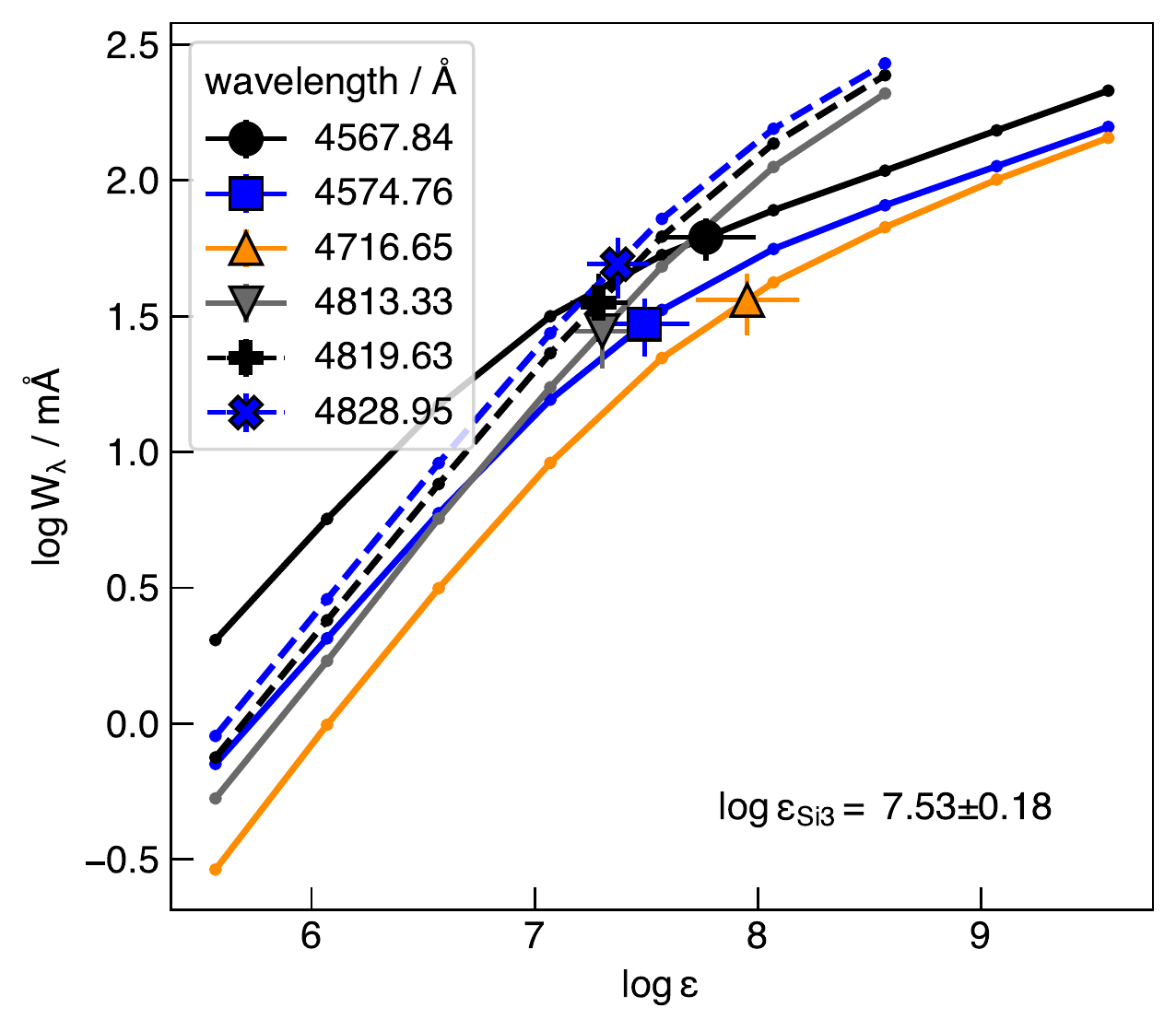}
    \includegraphics[width=\columnwidth]{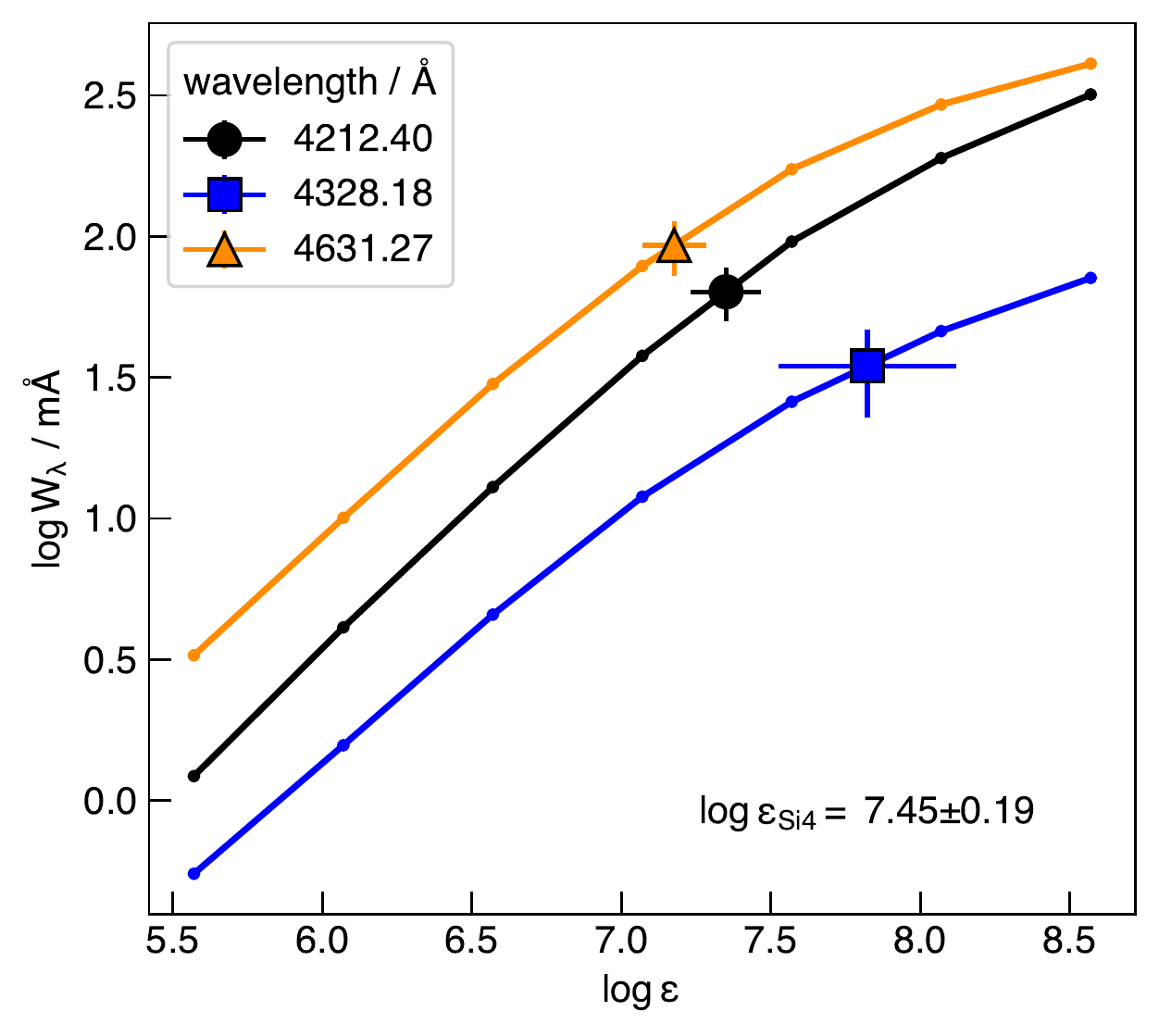} \\
    \includegraphics[width=\columnwidth]{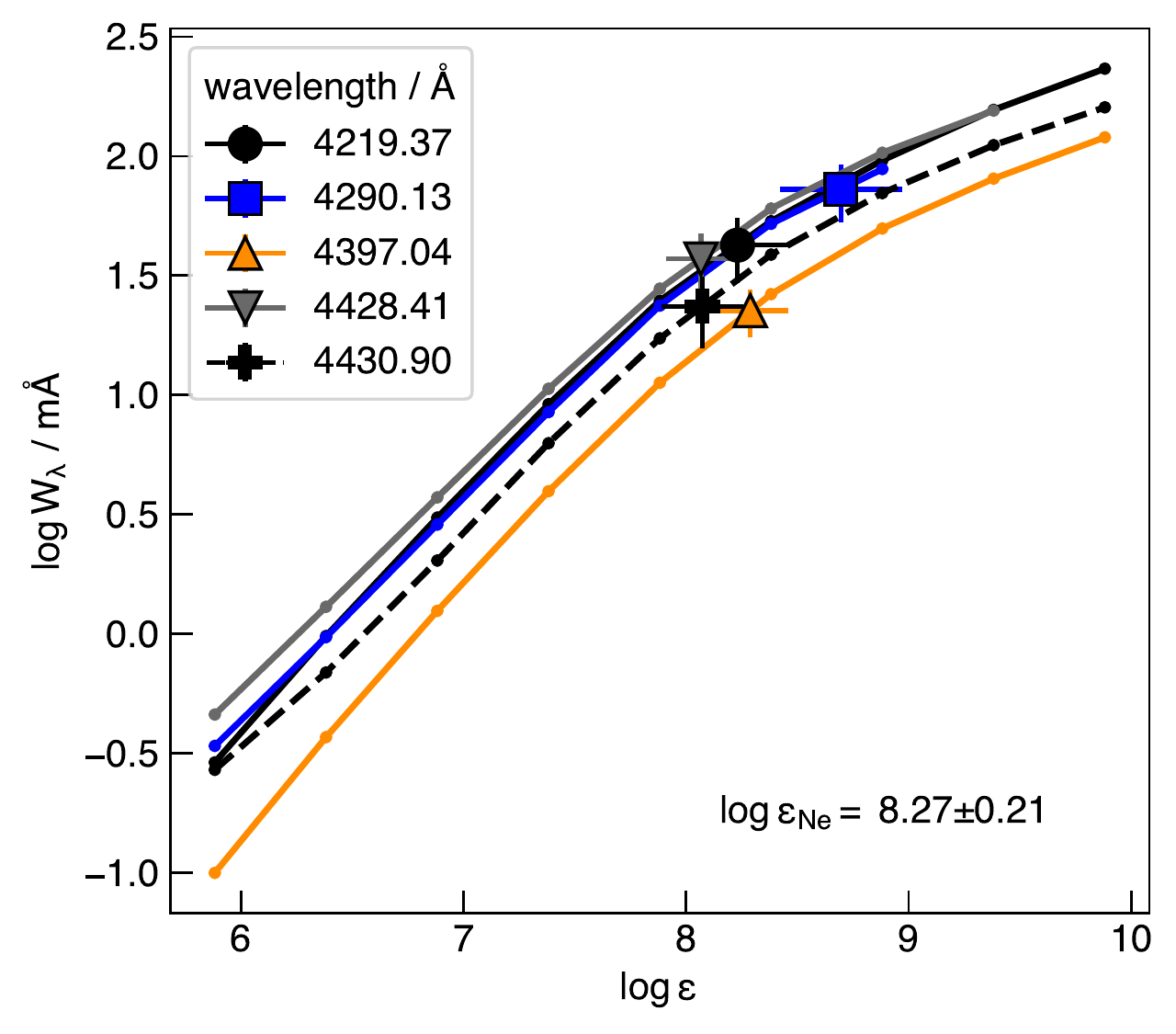}
    \includegraphics[width=\columnwidth]{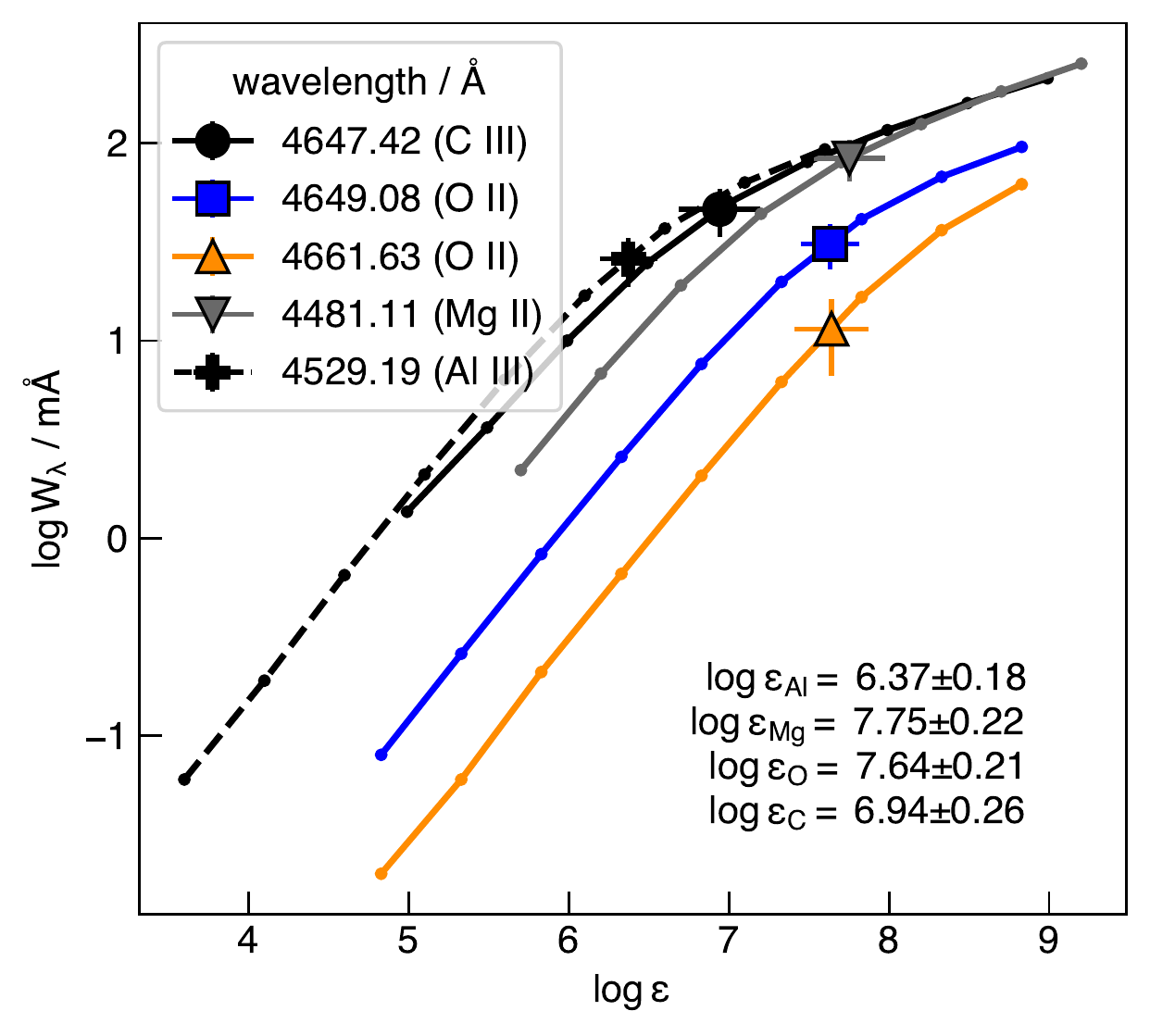}
    \caption{Equivalent width measurements plotted onto model curves of growth, along with corresponding abundances for each ion.}
    \label{fig:curves}
\end{figure*}

\begin{figure*}
	\includegraphics[width=0.9\textwidth]{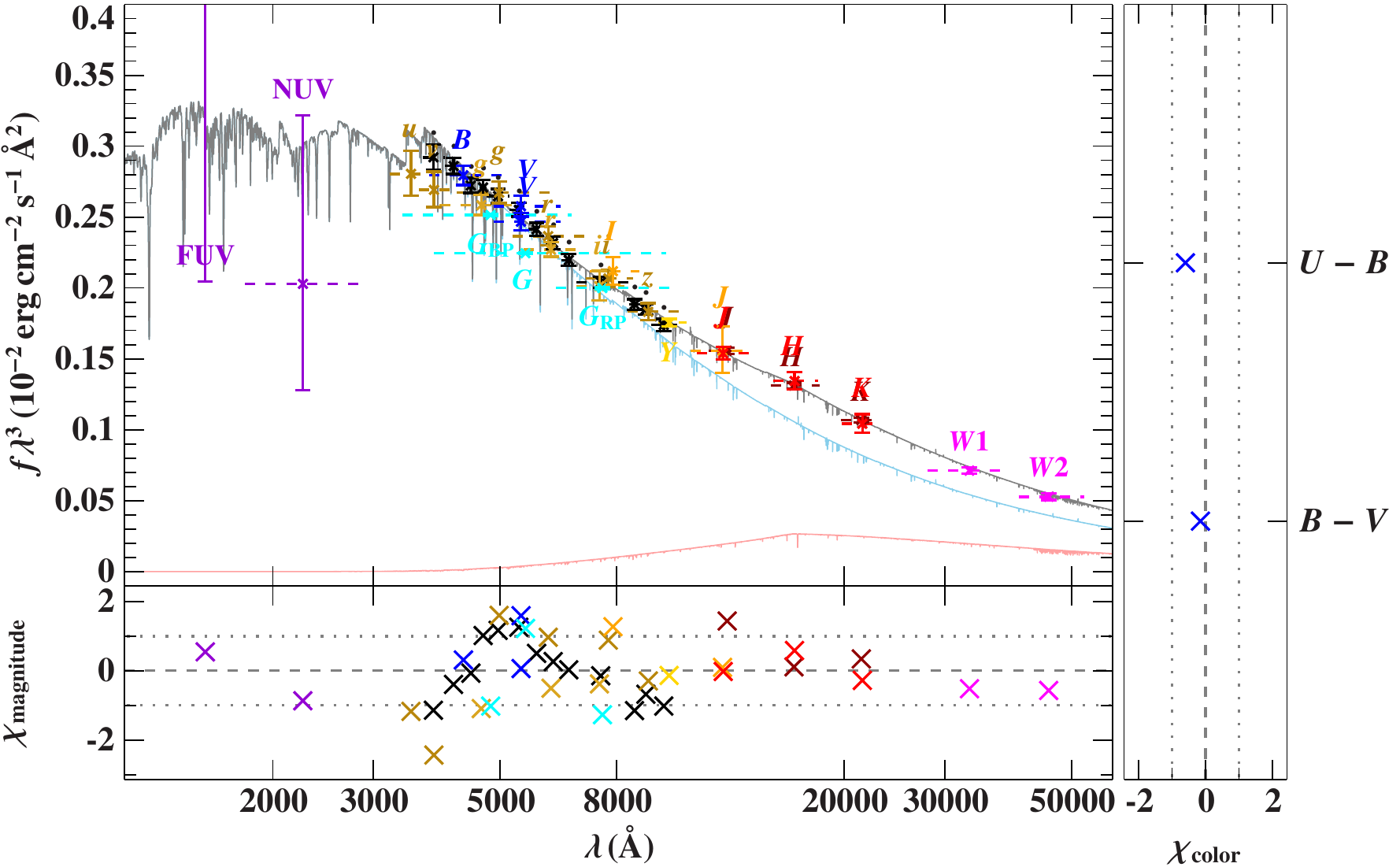}
        \includegraphics[width=0.9\textwidth]{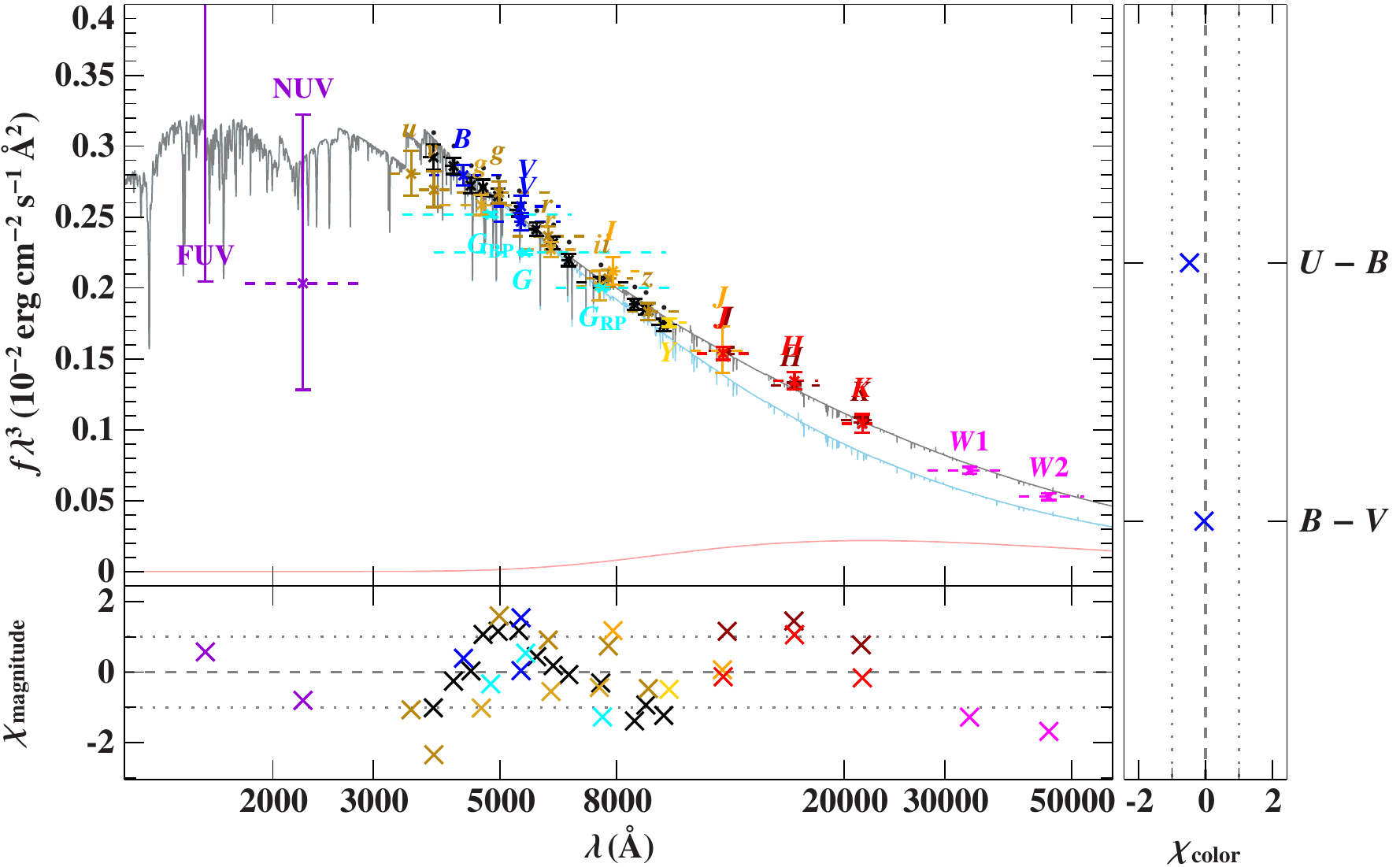}
    \caption{Spectral energy distribution fitted using {\sc isis}. The primary component flux is in blue whereas the secondary is in red. The added flux of both components is in grey. The upper plot shows the stellar fit whereas the lower plot shows the blackbody fit. Residuals are also given for U-B and B-V colours on the right-hand side of the plots and for flux on the bottom of the plots. The origin of the photometric data are: GALEX \citep[FUV and NUV, purple;][]{bianchi11}, SkyMapper \citep[u, v, g, r, i and z, olive;][]{onken19},
    APASS \citep{Henden2016}, 
    \textit{Gaia} \citep[XP in black, G, GRP and GBP in cyan;][]{gaia22}, Johnson \citep[B and V, blue;][]{Henden2016,O'Donoghue2013}, SDSS \citep[g, r, i, yellow;][]{Henden2016}
    DENIS \citep[I and J, orange;][]{denis03}, DES \citep[Y, yellow;][]{abbott22}, 2MASS \citep[J, H and K, red;][]{cutri03}, VISTA \citep[J, H and K, dark red;][]{cross12} and WISE \citep[W1 and W2, magenta;][]{schlafly20}.}
    \label{fig:sed}
\end{figure*}

\subsection{Abundances}
\label{sec:abunds}
The HRS spectrum was used to measure the equivalent widths of metal lines. Lines were included if there were no strong blends and if the depth exceeded $2\sigma$, where $\sigma$ is the noise in the continuum. The equivalent widths are listed in Table~\ref{tab:eqw}. These equivalent widths were used to estimate abundances using the {\sc spectrum} spectral synthesis code and the {\sc sterne} atmosphere modelling code. The {\sc sterne} model was calculated at $T_{\rm eff}=40$\,kK, $\log{g/\mathrm{cm\,s^{-2}}}$ and $n_{\rm He}=0.95$. These LTE abundances were used to create the model grid described in Sec.~\ref{sec:atm_params} and are listed under the $\log{\epsilon_{i,{\rm LTE}}}$ column of Table~\ref{tab:my_abn}.

To perform the non-LTE abundance analysis, a {\sc tlusty} model with the parameters obtained from the {\sc tlusty}/{\sc synspec} model grid fit to the RSS data was computed. As in the model grid, the abundances were taken from the {\sc sterne}/{\sc spectrum} LTE analysis. H, He and all metals listed in Table~\ref{tab:my_abn} were calculated in non-LTE.

Using this model, {\sc synspec} spectra were computed varying the abundance of individual metals. Nine model spectra were computed for each metal, where the logarithmic abundance varied between $\epsilon_0-2$ and $\epsilon_0+2$ in steps of 0.5, where $\epsilon_0$ is the original logarithmic abundance. The equivalent widths of the metal lines were then measured in the model spectra at each abundance step (sometimes excluding the lowest abundances where the line was too weak to measure or the highest abundances where an emission core developed).

Combining the model curves of growth for each metal line with the equivalent widths from the HRS data (Table~\ref{tab:eqw}), an abundance for the corresponding ion was measured from each line. These were averaged to infer an abundance for the metal. In the case of N and Si, lines from both of the ions present in the spectrum were used together to give the abundance of the metal.

These measurements were done twice, with the second pass using the atmospheric structure of a {\sc tlusty} model with abundances updated according to the results of the first pass. The abundance of an element was updated if its first pass abundance differed from the initial abundance by more than 0.5 dex. Thus, the final model used $\log{\epsilon_{\rm C}}=6.99$, $\log{\epsilon_{\rm N}}=8.89$, $\log{\epsilon_{\rm{Mg}}}=7.80$ and $\log{\epsilon_{\rm{Si}}}=7.57$, where the element is indicated by the subscript. Other elements remained the same as in the first pass. The final abundances measured from the second pass are listed in the $\log{\epsilon_{i}}$ column of Table~\ref{tab:my_abn}. The corresponding curves of growth are plotted in Fig~\ref{fig:curves}, where the lines of the individual ions of N and Si have been separated for clarity.

\subsection{Spectral energy distribution}
\label{sec:sed}
Photometric observations were used to fit the spectral energy distribution (SED) of the star using the {\sc isis} fitting tool \citep{heber18,irrgang21}. This fit is shown in Fig.~\ref{fig:sed}. The spectrum used for the fit was computed using the abundance analysis {\sc tlusty} model (Table~\ref{tab:model_params}). The photometric data showed an excess of flux in the red end; thus, a second component was used for the SED fitting. This was done for two cases: a stellar component and a blackbody. The surface gravity of the secondary component in the stellar case is not constrained by the SED and so was fixed at $\log{g/{\mathrm{cm\,s}^{-2}}}=5.0$ (an estimate for a main sequence star of roughly the temperature obtained from the fit). The results of the SED fitting are given in Table~\ref{tab:sed}. For the blackbody fitting, a temperature of 4.2$^{+0.7}_{-0.6}$\,kK was found for the secondary component. The ratio of the surface area of the blackbody with respect to the subdwarf was found to be 4.2$^{+0.8}_{-0.7}$.

\begin{table}
	\centering
	\caption{Results from the photometric fitting of the SED. Columns labelled (1) are results from fitting the second component as a star, whereas the column labelled (2) fitted the secondary component as a blackbody. The values for radius and mass are the mode of the probability density whilst the uncertainties correspond to the highest density intervals for 68\% confidence.}
	\label{tab:sed}
	\begin{tabular}{lccc} 
		\hline
		 & Primary (1) & Primary (2) & Secondary (1) \\
		\hline
		$T_{\rm{eff}}$ / kK & 40.78 (fixed) & 40.78 (fixed) & 4.8$^{+0.8}_{-0.7}$ \\
		$\log{g/{\mathrm{cm\,s}^{-2}}}$ & 5.90 (fixed) & 5.90 (fixed) & 5.0 (fixed) \\
		Radius / R$_{\odot}$ & 0.139$\pm0.005$ & 0.141$\pm0.005$ & 0.250$^{+0.024}_{-0.015}$ \\
		Mass / M$_{\odot}$ & 0.44$^{+0.32}_{-0.19}$ & 0.46$^{+0.31}_{-0.19}$ & 0.226$^{+0.045}_{-0.027}$ \\
		\hline
	\end{tabular}
\end{table}

\begin{figure*}
	\includegraphics[width=\linewidth]{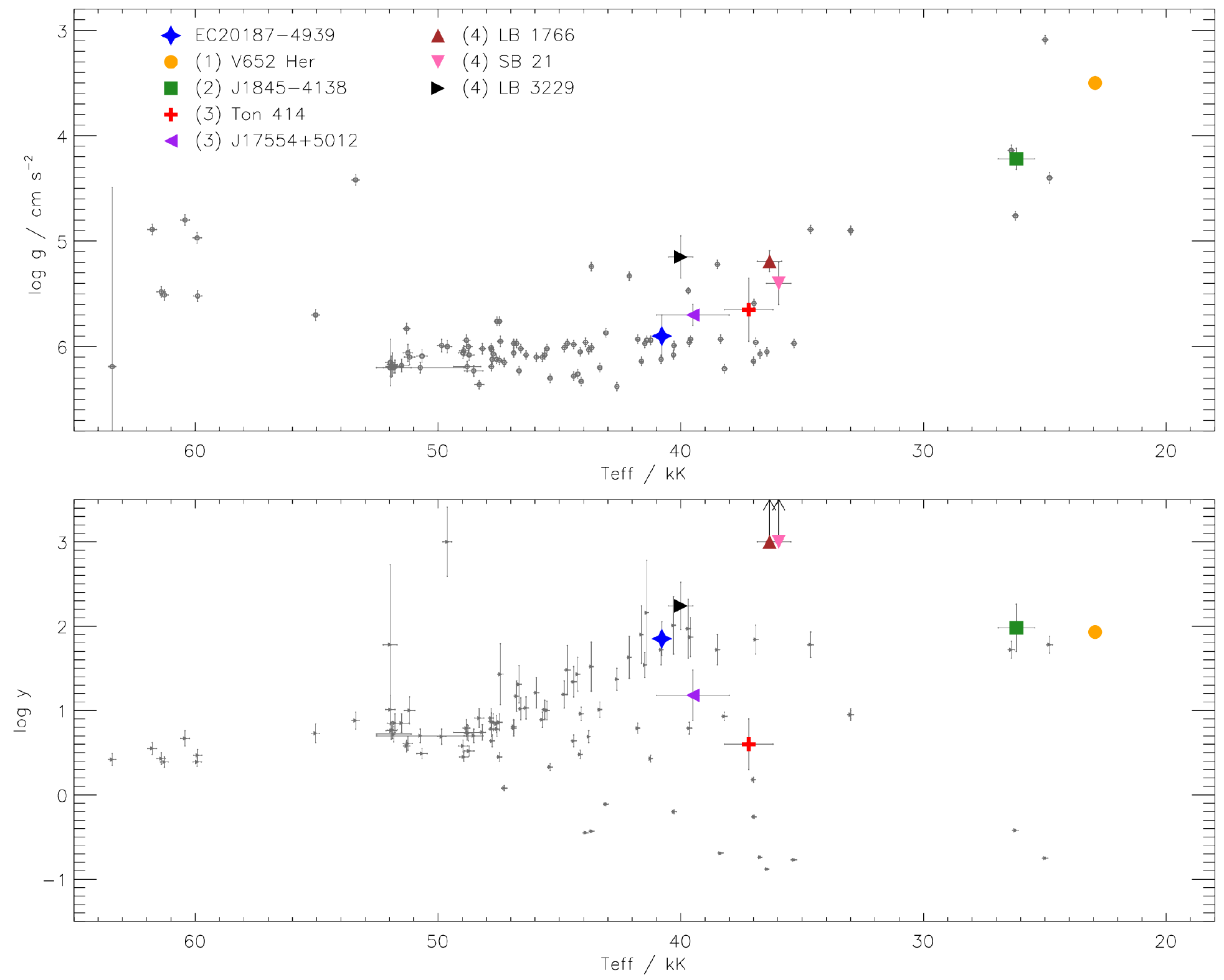}
    \caption{Position of \mystar{} in $\log{g}$-$T_{\rm eff}$ and $\log{y}$-$T_{\rm eff}$ space along with other stars from the \citet{jeffery21a} SALT survey. Other He-rich hot subdwarfs and extreme helium stars have been labelled; these are the same stars with sub-solar C abundance which are plotted later in Fig.~\ref{fig:abund_plot}. Of these, only \mystar{}, J1845-4138 and LB 3229 are included in the SALT survey. The source of the data for the labelled stars has been given in the legend and corresponds to (1) \citet{jeffery01b}, (2) \citet{jeffery17b}, (3) \citet{naslim20} and (4) \citet{naslim10}.}
    \label{fig:gty}
\end{figure*}

\begin{figure*}
    \centering
    \includegraphics[width=\columnwidth]{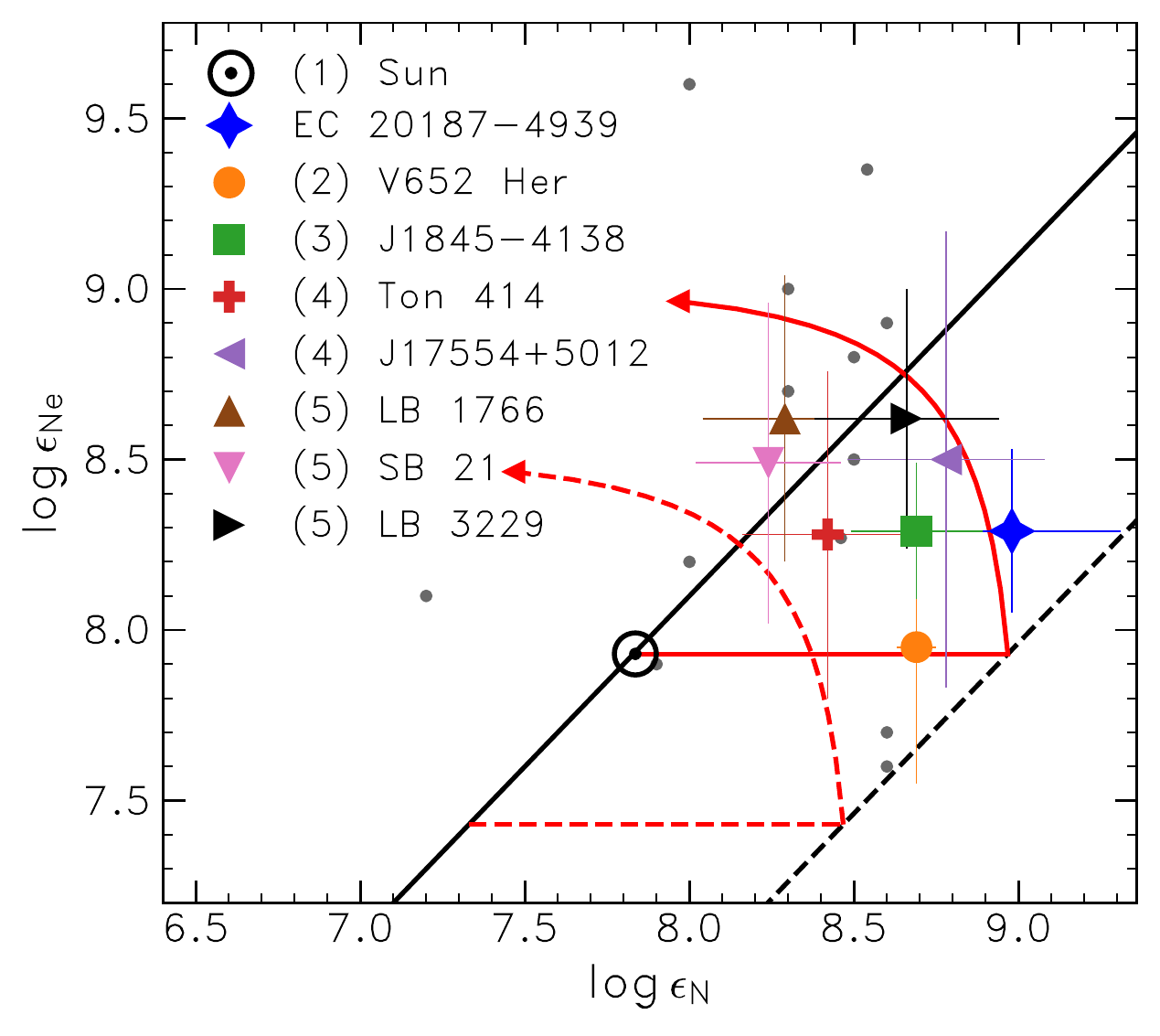} \includegraphics[width=\columnwidth]{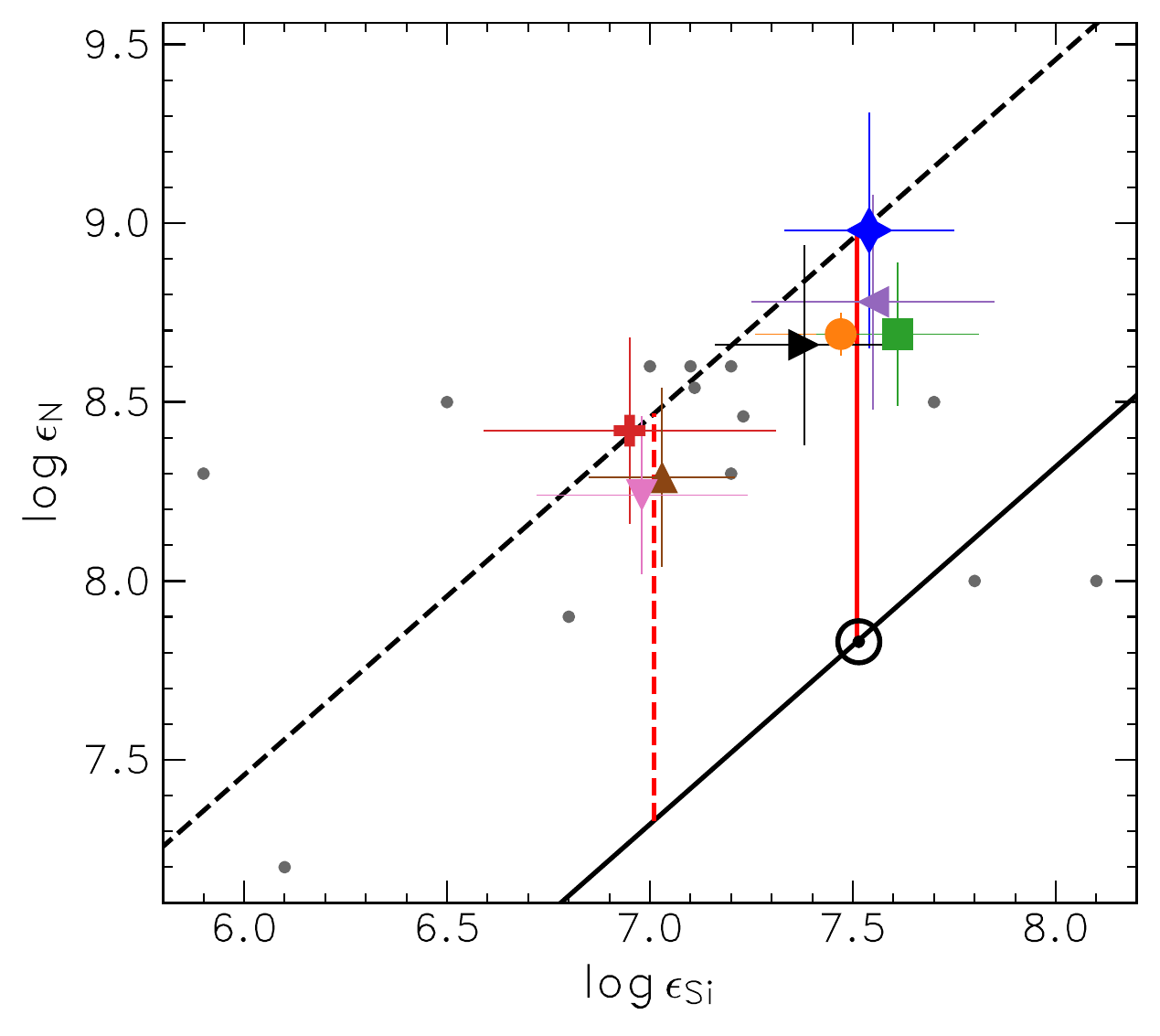} \\
        \caption{Ne versus N abundance and N versus Si abundance for \mystar{} and other hot subdwarfs and extreme helium stars. The abundances  are sourced from the papers labelled 1 to 5 at the end of this caption. Stars with sub-solar C abundances have been labelled and given a coloured marker, whereas stars with super-solar C are shown as grey dots (these are mostly extreme helium stars from \citet{jeffery11a}, with one hot subdwarf from \citet{naslim10}). The Sun, along with a solid black line representing scaled solar abundances, is also plotted. The dashed black line shows the scaled solar abundance but with all C and O converted to N. The solid red arrow on the left-hand plot shows the path of a star in the diagram that begins with solar abundances, undergoes CNO processing and then forms $^{22}$Ne by alpha captures onto $^{14}$N. The dashed red line shows the same path starting at $10^{-0.5}$ times solar abundance. The equivalent paths are shown on right-hand plot as vertical movement in the N-Si abundance plane. Sources for the abundance data are indicated by number in the legend, and are listed here: (1) \citet{asplund09}, (2) \citet{jeffery01b}, (3) \citet{jeffery17b}, (4) \citet{naslim20}, (5) \citet{naslim10}.}
    \label{fig:abund_plot}
\end{figure*}

\section{Discussion}
\label{sec:disc}
\subsection{Atmospheric Parameters}
\label{sec:disc_atmos}
\mystar{} has been included as part of a wider survey of hot subdwarfs \citep{jeffery21a}. This previous analysis differed from the current work by using an LTE grid for the fitting of $T_{\rm eff}$, $\log{g}$ and $n_{\rm He}$, with a slightly coarser grid spacing in $T_{\rm eff}$ and $\log{g}$ than in the current work. We departed from the methodology of \citet{jeffery21a} by choosing to use the N ion balance to measure the temperature, which was then held fixed in the chi-squared minimisation fitting, and by using non-LTE atmosphere models. Nevertheless, our results for $T_{\rm eff}$ and $\log{g}$ are in agreement with \citet{jeffery21a}, although we find a slightly higher He abundance.

Figure\,\ref{fig:gty} shows \mystar{} with the other stars in the \citet{jeffery21a} survey in the $\log{g}-T_{\rm eff}$ and $\log{y}-T_{\rm eff}$ planes. Its position in $\log{g}-T_{\rm eff}$ is not unusual, as it lies at the cool end of the cluster of sdO stars at $\log{g}\sim6$. In $\log{y}-T_{\rm eff}$, it is one of the more H-deficient (high $\log{y}$) stars present.

\subsection{Abundances}
\label{sec:disc_abunds}
The metal composition of \mystar{} is characterised by high N and low C and O, which is consistent with material processed by the CNO cycle. \mystar{} therefore fits into the He-strong, carbon-weak category in the Drilling classification system \citep{drilling13}. Ne is slightly enriched compared to solar, whereas other metal abundances are similar to solar.

To better understand the abundance pattern of \mystar{} and its relation to similar stars, Fig~\ref{fig:abund_plot} shows its abundances of N, Ne and Si, which is used as a proxy for metallicity, along with those of other helium-rich hot subdwarfs and extreme helium stars for which these abundances have been measured. The solid black line represents scaled solar abundances. Since \mystar{} is N-enriched, a dashed black line has been added to show the location of scaled solar abundance when the entire mass of C and O has been converted into N, as an approximation for the results of nuclear processing by the CNO cycle.

The abundances of \mystar{} can be used to rule out certain formation processes. The hot flasher route, where a post-red giant branch star undergoes a He core flash and loses mass, can produce He-rich subdwarfs but does not appear to be consistent with the low C/N ratio ($\sim0.01$) seen in \mystar{}. This is because the He core flash triggers convection which mixes C from the core to the surface of the star, leading to much higher C/N ratios of between $\sim9.5$ to $\sim0.85$ \citep{bertolami08}.

White dwarf mergers are a likely formation channel for hot subdwarfs and nuclear processing such as alpha captures are expected to occur in the merger process \citep{zhang12a}. The $^{14}$N($\alpha,\gamma$)$^{18}$O and $^{18}$O($\alpha,\gamma$)$^{22}$Ne reactions present a route to form Ne from N- and He-rich material. The solid red line on the left-hand plot of Fig.~\ref{fig:abund_plot} shows the path from solar metal abundances to CNO-processed (horizontal line) and then a locus of progressively more conversion of $^{14}$N to $^{22}$Ne (curved line). This curved line extends from 0\% to 90\% conversion of $^{14}$N to $^{22}$Ne. An equivalent path starting at $10^{-0.5}$ times solar abundances is shown in dashed red, showing how a change in metallicity translates the path in the abundance plane.

The right-hand plot of Fig~\ref{fig:abund_plot} shows N abundance versus Si abundance for the same stars. The vertical red lines are equivalent to the red paths in the left-hand plot; CNO processing moves the abundances upwards from the solid black  to the dashed black line, then alpha captures onto $^{14}$N move the abundances vertically downwards again.
Assuming the Si abundance and the initial N abundance scale similarly with overall metallicity, then stars with lower Si abundance (e.g. Ton 414, SB 21 and LB 1766) would lie along the dashed red line on the left-hand plot whereas stars with solar Si abundance would lie along the solid red line.

The very high N abundance of \mystar{} does not appear to be associated with a high metallicity, since its Si and Al abundances are roughly solar. If CNO cycle processing followed by alpha captures on $^{14}$N is indeed the main route by which the N abundance in these stars is altered during evolution, then this indicates that either the merger was slow and cool enough that alpha captures were not frequent during formation or that this processed material was not mixed to the surface of the resultant star.

\citet{zhang12a} produced models of He white dwarf mergers which can explain the He- and N-rich surface composition of \mystar{}. They proposed three formation processes: the slow, fast and composite mergers. In the slow merger case, the secondary star is disrupted into a cold disk which then slowly accretes onto the primary (over several million years). In the fast merger case, the secondary forms a hot corona around the primary after transferring its entire mass to the primary within a few minutes. The composite case is a mixture of the two, where only some material from the secondary forms a disk whilst the rest is contained in the hot corona. In all cases, triple-$\alpha$ burning is expected to ignite as the material from the secondary collides with the surface of the primary. However, only in the fast merger and higher-mass composite mergers is the resultant $^{12}$C mixed to the surface of the final merger product by convection. Thus, a low-C surface composition, as seen in \mystar{}, is consistent with a slow merger or low-mass ($<0.65$\,M$_{\odot}$) composite merger. $\alpha$ capture products such as $^{18}$O and $^{22}$Ne are also less abundant in these cases, again consistent with the low O abundance and only slightly enriched Ne abundance of \mystar{}.

Considering the lack of significant rotational broadening in the spectrum, \mystar{} does not appear to be a fast rotator. This does not necessarily mean that it is not the product of a binary merger, as long as angular momentum can be dissipated sufficiently during the merging process. This is certainly possible through the formation of a viscous disk \citep{lyndenbell74}, which will deposit a small fraction of the total angular momentum onto the central star. The case of collisionless disks (which are less efficient in dissipating angular momentum) was studied by \citet{gourgouliatos06}, who concluded that some angular momentum must be lost during the double He white dwarf merger, otherwise the fully contracted end product would be rotating supercritically. \citet{schwab18} explored mass loss during the short post-merger giant phase as a mechanism for angular momentum dissipation, finding that three different mass loss schemes yielded slowly-rotating end products. Additional angular momentum may be lost through magneto-dipole radiation shortly after the merger \citep{garciaberro12}. Therefore, the slow rotation of \mystar{} does not exclude a double white dwarf merger as a formation route.

Placing \mystar{} in the context of other carbon-weak extreme helium stars and subdwarfs is difficult because of the small numbers available for comparison, and the wide range of surface hydrogen abundances observed within the sample. Inferring an evolutionary connection between \mystar{} and, say, V652\,Her, for which the double white-dwarf merger model is a strong contender \citet{saio00},  or J1845--4138, is tempting but requires confirmation via identification of additional stars with similar surface chemistry and mass.

\subsection{Spectral energy distribution}
\label{sec:disc_sed}
The SED fitting allowed a mass and radius to be found for \mystar{} (Table~\ref{tab:sed}). The masses of $0.44^{+0.32}_{-0.19}$\,M$_\odot$ (stellar secondary component) and $0.46^{+0.31}_{-0.19}$\,M$_\odot$ (blackbody secondary component) fall within the range ($<0.65$\,M$_{\odot}$) predicted by \citet{zhang12a} for the production of He-rich, C-poor hot subdwarfs by the composite merger route. Of the two secondary components used in the SED fitting (star and blackbody), the best fit was achieved with the star (see the flux residuals in Fig.~\ref{fig:sed}). However, the mass of the secondary component in this case is very small ($0.226^{+0.045}_{-0.027}$\,M$_{\odot}$) considering its temperature of 4.8$^{+0.8}_{-0.7}$ kK. For the SED fit in Sec.~\ref{sec:analysis}, the secondary component was assumed to have $\log{g/{\rm cm\,s^{-2}}}=5$, which is on the upper end for a main sequence star at this temperature. The derived mass of $0.226^{+0.045}_{-0.027}$\,M$_{\odot}$ is therefore effectively an upper limit (a lower assumed surface gravity would produce a lower mass). For comparison, the expected mass for a main sequence star at $\sim4.8$\,kK is $\sim0.7$\,M$_{\odot}$ \citep[using MESA Isochrones \& Stellar Tracks;][]{choi16}. Thus, the excess red flux seen in the photometric data is unlikely to be a main sequence stellar companion.

The cause of the excess is possibly a contaminating infrared source, however there are no other {\it Gaia} sources in the vicinity. Except in the case of a chance alignment within less than 1'', this excludes a blend.

In the case of treating the secondary source as a blackbody, the SED fitting found an effective surface area ratio of 4.2$^{+0.8}_{-0.7}$ between the blackbody and the subdwarf.
By treating the blackbody source as a circular disk, its radius, $R_{\rm bb}$, can be related to the radius of the subdwarf, $R_{*}$, as $(R_{\rm bb}\sin{i})^2/R_{*}^2=4.2$, where $i$ is the inclination angle. As an example, $R_{\rm bb}\approx2R_*$ for the range $0^{\circ}<i<60^{\circ}$. Taking the subdwarf radius of 0.141\,R$_{\odot}$ from the SED fit, this corresponds to $\approx0.3$\,R$_{\odot}$. It is possible that this relatively small size is consistent with the residue of a white dwarf merger. Whether the blackbody temperature of 4.2$^{+0.7}_{-0.6}$\,kK is consistent depends on how recent the merger was, since the residue would likely cool and disperse after the event.

\section{Conclusions}
\label{sec:conclusions}
We have analysed \mystar{} using non-LTE {\sc tlusty/synspec} models. We have re-measured the temperature, surface gravity and helium abundance, finding good agreement with the previously published results in \citet{jeffery21a}. We now have metal abundances for the star, allowing it to be classified within the C-poor group of He-rich hot subdwarfs. The N abundance is particularly high and the Ne abundance provides a third key diagnostic.

Publicly available photometric data have been used to fit the SED and show the presence of an excess of flux at the red end. The origin of this extra flux does not appear to be explained by a stellar source due to inconsistencies between the fitted mass and temperature of the secondary component. There are no nearby stellar sources visible in photometric data which could have contributed to the infrared excess. A merger residue is potentially consistent, although this depends on how recent the merger event was. In any case, whether the secondary component is fitted as a star or a blackbody does not affect the fit of the primary component.

Whilst the sample size is too small to demonstrate an evolutionary connection between \mystar{} and carbon-weak extreme helium stars and hot subdwarfs unequivocally, the low carbon and high helium abundances of \mystar{}, along with the mass of 0.44$^{+0.32}_{-0.19}$\,M$_{\odot}$, are consistent with formation via a merger of low-mass helium white dwarfs \citep{zhang12a}. Mildly enhanced neon compared to solar also suggests alpha captures on $^{14}$N occurred during the merger. Metal abundance measurements of more hot subdwarfs would be valuable to identify trends with helium abundance, temperature and surface gravity, which could allow evolutionary connections to be established more firmly.

\section*{Acknowledgements}

The authors are indebted to the UK Science and Technology Facilities Council via UKRI Grant No. ST/V000438/1 for grant support. The Northern Ireland Department for Communities funds the Armagh Observatory and Planetarium (AOP), providing for AOP membership of the United Kingdom SALT consortium (UKSC). The data for \mystar{} were obtained through the UKSC. Part of this investigation was carried out with financial support from the Trinity College Dublin School of Physics.

\section*{Data Availability}

The models and observations used in this investigation will be made available upon reasonable request to the authors.


\bibliographystyle{mnras}
\bibliography{trinity_star}







\bsp	
\label{lastpage}
\end{document}